\documentclass[10pt, conference, compsocconf]{IEEEtran}
\usepackage{amssymb, amsmath, amsthm}
\usepackage{graphics, color}
\usepackage{multirow}

\newcommand{\Rmnum}[1]{\MakeUppercase{\romannumeral #1}}
\usepackage{tikz}
\ifCLASSINFOpdf
\else
\fi
\begin{document}
%
\title{A Task-Type-Based Algorithm for the Energy-Aware Profit Maximizing Scheduling
Problem in  Heterogeneous computing systems}


\author{\IEEEauthorblockN{Weidong Li \thanks{Correspondence: weidong@ynu.edu.cn (W. Li), xjzhang@ynu.edu.cn (X. Zhang)}, Xi Liu, Xuejie Zhang, Xiaobo Cai}
\IEEEauthorblockA{Yunnan University\\
Kunming, China\\
Email: \{weidong, xliu, xjzhang, xbcai\}@ynu.edu.cn}
}


%


\maketitle

\begin{abstract}
In this paper, we design an efficient algorithm for the energy-aware
profit maximizing scheduling problem, where the high performance
computing system administrator is to maximize the profit per unit
time. The running time of the proposed algorithm is depending on the
number of task types, while the running time of the previous
algorithm is depending on the number of tasks. Moreover, we prove
that the worst-case performance ratio is close to 2, which maybe the
best result. Simulation experiments show that the proposed algorithm
is more accurate than the previous method.
\end{abstract}


%
\IEEEpeerreviewmaketitle

\section{Introduction}

\subsection{Background and Motivation}
In high-performance computing (HPC) systems, it is well known that
when the performance is increased, the power consumption is
increased, as well as the electricity costs for the operators are
increased. Recently, the high cost of the HPC systems has lead to
research that designs an efficient resource allocation algorithm to
reduce the required energy consumption \cite{Tarplee14}. By
combining the energy and performance objectives into a single profit
objective,  Tarplee et al. \cite{Tarplee14} introduced a novel
monetary-based model for HPC where there is a financial distinction
between the service provider and the users.  In HPC systems, there
are two important facts: (a) The HPC systems are often composed of
different types of machines; (b) There are a large number of tasks
but only small number of task types. By solving a linear program and
rounding carefully, they \cite{Tarplee14} designed an efficient
algorithm to find a feasible schedule.

In \cite{Tarplee14},  a lower bound on the finishing times of a
machine type is used to replace makespan, which is defined as the
maximum finishing time of all machines.  Therefore, the  proposed
 mathematical  model is inaccurate. For the
proposed algorithm  \cite{Tarplee14}, in the rounding process, the
energy consumption maybe increased, which can be avoided by using a
different method. Moreover, the running time is depending on the
number of tasks, which can be improved, too. Most importantly, the
worst-case performance ratio of the proposed algorithm
\cite{Tarplee14} is not given.

\subsection{Contributions and Outline}
This paper presents an accurate mathematical model for the problem
proposed in \cite{Tarplee14}. A polynomial-time algorithm is then
developed to find a feasible solution for the proposed model.

The contributions of this paper are:

1) An accurate mathematical model;

2) A task-type-based  algorithm to find a more accurate feasible
solution, whose running time is independent of the number of tasks;

3) The worst-case performance ratio.

The remainder of this paper is organized as follows. The next
section proposes the accurate mathematical model. Section
\uppercase\expandafter{\romannumeral3} presents the task-type-based
algorithm and proves the worst-case performance ratio. Section
\uppercase\expandafter{\romannumeral4} gives the experimental
results. The last section discusses the useful extensions to the
proposed model and lists ideas for future work.

\section{The Mathematical Model}
As in \cite{Tarplee14}, a user submits a bag-of-tasks to process,
where each task is indivisible and independent of all the other
tasks. The cost to the organization for processing a bag-of-tasks is
the cost of electricity. The organization or service provider should
maximum the profit per bag, which is equal to the price minus the
cost. However, the bag-of-tasks can take a considerable amount of
time to compute when trying to increase the profit by reducing
electricity costs. Thus, it is more reasonable for an organization
to maximize the profit per unit time.

Formally, assume that there are $T$ task types and $M$ machine
types. Let ${\cal T}_i$ be the set of tasks of type $i$ and $T_i$ be
the number of tasks in ${\cal T}_i$. Similarly, let ${\cal M}_j$ be
the set of machines of type $j$ and $M_j$ be the number of machines
in ${\cal M}_j$. Denote by $x_{ij}$ the number of tasks of type $i$
assigned to a machine of type $j$, where $x_{ij}$ is the primary
decision variable in the optimization problem. As the definitions
frequently used in scheduling algorithms \cite{Tarplee14}, let
$\textbf{ETC}$ be a $T\times M$ matrix where $ETC_{ij}$ is the {\it
estimated time to compute} for a task $i$ on a machine $j$.
Similarly, let $\textbf{APC}$ be a $T\times M$ matrix where
$APC_{ij}$ is the {\it average power consumption} for a task $i$ on
a machine $j$.

Since tasks are indivisible in most cases, the $x_{ij}$ tasks of
type $i$ may not be allocated equally to the $M_j$ machines of type
$j$. For every machine $j_k\in {\cal M}_j$, let $x_{ijk}$ be the
number of tasks of type $i$ assigned to machine $j_k$. Clearly,
$x_{ij}=\sum_{k:k\in {\cal M}_j}x_{ijk}$. The finishing time of a
machine
 $j_k\in {\cal M}_j$, denoted by $F_{jk}$, is given by
 \begin{eqnarray}
F_{jk}=\sum_{i=1}^{T}x_{ijk}ETC_{ij}.
\end{eqnarray}
Thus, the maximum finishing time of all machines (i.e.,
\emph{makespan}), denoted by $MS({\mathbf x })$, is given by
 \begin{eqnarray}
MS({\mathbf x })=\max_j \max_{k: j_k\in {\cal M}_j}F_{jk}.
\end{eqnarray}
In this paper, for convenience, machines are turned off when not
use, which means that the energy consumed by the bag-of-tasks is
given by:
\begin{eqnarray}
 E({\mathbf
x
})=\sum_{j=1}^{M}\sum_{i=1}^{T}x_{ij}APC_{ij}ETC_{ij}.\end{eqnarray}

Let $p$ be the price customer pays and $c$ be the cost per unit of
electrical energy. The profit that the organization receives by
executing a bag-of-tasks is $p-c E({\mathbf x })$. The {\it
Energy-Aware Profit Maximizing Scheduling} (EAPMS) Problem defined
in \cite{Tarplee14} attempting to maximize the profit per unit time
can be formulated as the following nonlinear integer program (NLIP):
 \begin{eqnarray}
 \begin{split}
 Maximize_{\mathbf{x}}  \hspace{2mm} &\frac{p-cE({\mathbf x })}{MS(\mathbf{x})}&\\
\text{subject to }    &\forall i \hspace{2mm}  \sum_{j=1}^{M}\sum_{k: j_k\in {\cal M}_j}x_{ijk}=\sum_{j=1}^{M}x_{ij}=T_i;& \\
  &\forall j  \hspace{2mm} F_{jk}\leq MS(\mathbf{x}), \text{for each }   j_k\in {\cal M}_j;&\\
     &\forall i, j  \hspace{2mm} x_{ijk}\in  \mathbb{Z}_{\geq 0},   \text{for each }  j_k\in {\cal M}_j.&
\end{split}\end{eqnarray}
The objective of (4) is to maximize the profit per unit time, where
$\mathbf{x}$ is the primary decision variable. The first constraint
ensures that all tasks of different types in the bag are assigned to
some machine type. Because the objective is to maximize the profit
per unit time, which is equivalent to minimize makespan, the second
constrain ensures that $MS(\mathbf{x})$ is equal to the maximum
finishing time of all machines.

 \section{A Task-Type-Based Algorithm}
\subsection{Overview}
¡¡ Note that (4) is a nonlinear integer program, which can not be
solved optimally in polynomial time. To obtain an approximate
solution of (4), one possible way is to convert (4) to an equivalent
linear program (LP), and then to round the optimal fraction solution
of LP to a feasible solution for  (4). In \cite{Tarplee14},
 the authors obtained a linear program using variable
substitution $r\leftarrow 1/MS_{LB}$ and $z_{ij}\leftarrow
x_{ij}/MS_{LB}$, where $MS_{LB}=\max_j
\frac{1}{M_j}\sum_{i=1}^{T}x_{ij}ETC_{ij}$ is a lower bound on the
makespan obtained by allowing tasks to be divided among all
machines. However, the approximation of this method would be bad
when the objective value is close to 0 or little tasks of type $i$
with large $ETC_{ij}$ are assigned to machines of type $j$. A
similar phenomenon is also observed by Tarplee et al.
\cite{Tarplee13}.

 To overcome the obstacle  mentioned above, we will use
a different method. We replace $MS(\mathbf{x})$ with a constant
$MS$, and then obtain an approximate  integer linear program  (ILP)
for (4). By rounding the optimal fraction solution for the
relaxation of ILP based on the classic rounding algorithm for
 the generalized assignment problem \cite{Shmoys93}, we obtain a feasible solution
for (4). It is desired to point out that, in our method, the tasks
of type $i$ such that $ETC_{ij}>MS$ will not be assigned to machines
of type $j$, which is to avoid increasing the makespan too much when
rounding the optimal fraction solution.

Let $LB$ be the optimal makespan by ignoring the energy consumption,
and $UB$ be the makespan of the feasible schedule by assigning each
task to the machine with minimum average power consumption
$APC_{ij}$. For any given constant $\epsilon>0$, Clearly, the
makespan $MS({\mathbf{x}^*})$ of the optimal solution $\mathbf{x}^*$
for (4)
 lies in  $[LB(1+\epsilon)^t, LB(1+\epsilon)^{t+1}]$, for some $t\in \{1,2,\ldots,\lceil\log_{(1+\epsilon)}
 UB/LB\rceil\}$.
By trying all possible values,  we will find a
 feasible makespan $MS$ such that
 $MS({\mathbf{x}^*})\in [MS/(1+\epsilon), MS]$,  where $MS=LB(1+\epsilon)^t$ for some $t$.
  For convenience, from now on, assume that
  $MS$ is a known constant satisfying
 \begin{eqnarray}
MS({\mathbf{x}^*})\leq MS\leq (1+\epsilon) MS({\mathbf{x}^*}).
 \end{eqnarray}

For a constant $MS$, as in \cite{Tarplee14}, our algorithm is
decomposed into two phases. This first phase rounds the fraction
optimal solution to obtain a schedule where the numbers $x_{ij}$ of
tasks of type $i$ assigned to machines of type $j$ are given. The
second phase assigns tasks to actual machines to produce the full
task allocation $x_{ijk}$. The next two subsections describe the two
phases of this recovery procedure in detail.

There are two main differences between Tarplee, Maciejewski, and
Siegel's (TMS, for short) method \cite{Tarplee14} and our
task-type-based (TTB, for short) method (depicted in Figure 1.): (1)
The TMS method uses one fractional solution to round while we use
multiple fractional solutions and choose the best one; (2) In the
first phase, the energy consumption may increase in Tarplee et al.'s
method while it will not increase in our method.
\subsection{$b$-Matching-Based Rounding}
 Note that if $ETC_{ij}>MS$,
the tasks of type $i$ can not be assigned to the machines of type
$j$ in the optimal solution, by the definition of $MS$. This implies
that $x_{ijk}=x_{ij}=0$, if $i,j,k$ satisfy that $ETC_{ij}>MS$ and
$j_k\in {\cal M}_j$.
 As mentioned in \cite{Tarplee14},
$\frac{1}{M_j}\sum_{i=1}^{T}x_{ij}ETC_{ij}$ is a lower bound on
$MS$. Since
\begin{center}
\begin{tikzpicture}[scale=1.3]
\draw[->](-1.1,0)--(1.6,0); \draw[->] [xshift=3.5cm]
(-1.1,0)--(1.6,0);
 \draw[->](-1.1,0)--(-1.1,1.5);
\draw[->] [xshift=3.5cm] (-1.1,0)--(-1.1,1.5); 

 \draw [thick](-1,1.4)--(-0.7,0.9)--(-0.2,0.6)--(0.5,0.3)--(0.9,0.2)--(1.4,0.1);
  \draw [xshift=3.5cm] [thick](-1,1.4)--(-0.7,0.9)--(-0.2,0.6)--(0.5,0.3)--(0.9,0.2)--(1.4,0.1);
¡¡ \filldraw [green] (-0.2,0.6) circle (0.3mm);
 \draw   (-0.2,0.6)--(0,0.7)--(0,0.9);
 \filldraw [gray] (0,0.7) circle (0.3mm);
  \filldraw [red] (0,0.9) circle (0.3mm); 

       \draw  [xshift=3.5cm] [thick](-0.7,0.9)--(-0.7,1.2);
   \filldraw [xshift=3.5cm] [green] (-0.7,0.9) circle (0.3mm);
    \filldraw [xshift=3.5cm] [gray] (-0.7,1.1) circle (0.3mm);
  \filldraw [xshift=3.5cm] [red] (-0.7,1.2) circle (0.3mm);

       \draw  [xshift=3.5cm] [thick](-0.45,0.75)--(-0.45,0.95);
   \filldraw [xshift=3.5cm] [green] (-0.45,0.75) circle (0.3mm);
    \filldraw [xshift=3.5cm] [gray] (-0.45,0.85) circle (0.3mm);
  \filldraw [xshift=3.5cm] [red] (-0.45,0.95) circle (0.3mm);

   \draw  [xshift=3.5cm] [thick](-0.2,0.6)--(-0.2,0.9);
   \filldraw [xshift=3.5cm] [green] (-0.2,0.6) circle (0.3mm);
    \filldraw [xshift=3.5cm] [gray] (-0.2,0.75) circle (0.3mm);
  \filldraw [xshift=3.5cm] [red] (-0.2,0.9) circle (0.3mm);

   \draw  [xshift=3.5cm] [thick](0.15,0.45)--(0.15,0.75);
   \filldraw [xshift=3.5cm] [green] (0.15,0.45) circle (0.3mm);
    \filldraw [xshift=3.5cm] [gray] (0.15,0.65) circle (0.3mm);
  \filldraw [xshift=3.5cm] [red] (0.15,0.75) circle (0.3mm);

    \draw  [xshift=3.5cm] [thick](0.5,0.3)--(0.5,0.6);
   \filldraw [xshift=3.5cm] [green] (0.5,0.3) circle (0.3mm);
    \filldraw [xshift=3.5cm] [gray] (0.5,0.5) circle (0.3mm);
  \filldraw [xshift=3.5cm] [red] (0.5,0.6) circle (0.3mm);

 \draw (1.2,-0.2) node {energy} [xshift=3.5cm] (1.2,-0.2) node { energy } ;
 \draw (-1.3,0.8)  node {\rotatebox{90}{\small makespan}} [xshift=3.5cm] (-1.3,0.8)  node {\rotatebox{90}{\small makespan}} ;
 \draw (0.3,-0.5) node {(a) TMS method} [xshift=3.5cm] node {(b) TTB method};
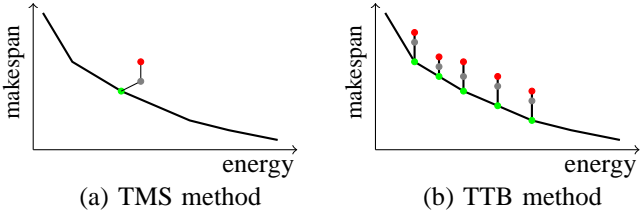
 \draw [xshift=2cm] (0,-1) node {\textbf{Figure 1.} Comparing the main ideas of two algorithms
};
\end{tikzpicture}
\end{center}
 \vspace{2mm}
$MS$ is constant close to $MS({\mathbf{x}^*})$, we can substitute
$MS$ for $MS(\mathbf{x})$ in  (4). Since $p,MS,c$ are constants, the
objective maximizing $(p-cE({\mathbf x }))/MS=p/MS-cE({\mathbf x
})/MS$ is equivalent to minimizing $ E({\mathbf x })$.
 Thus, we obtain an approximate equivalent integer programming
formula for NLIP (4):
\begin{eqnarray}
 \begin{split}
 Minimize_{\mathbf{x}}  \hspace{2mm} &E({\mathbf
x })= \sum_{j=1}^{M}\sum_{i=1}^{T}x_{ij}APC_{ij}ETC_{ij}&\\
\text{subject  to }    &\forall i \hspace{1mm} \sum_{j=1}^{M}x_{ij}=T_i;& \\
   &\forall j\hspace{2mm} \frac{1}{M_j}\sum_{i=1}^{T}x_{ij}ETC_{ij}\leq MS;&\\
   & x_{ij}\in  \mathbb{Z}_{\geq 0},   \text{for each }  i, j;&\\
     & x_{ij}=0,   \text{if }    ETC_{ij}>MS.&\\
\end{split}\end{eqnarray}

\vspace{1mm}\noindent {\bf Theorem~1.} {\em Any  optimal
solution  $\mathbf{x}^*$  for NLIP (4) is a feasible solution for (6).}   \\

 Replacing the constraint $x_{ij}\in \mathbb{Z}_{\geq 0}$ with
$x_{ij}>0$, we obtain the relaxation of  (6), which is a linear
program and can be solved in polynomial time. Noting that there are
$TM$ variables and $T+M$ nontrivial constraints, both are less than
that in the linear program (10) in \cite{Tarplee14}. By modifying
Shmoys \& Tardos's rounding method in \cite{Shmoys93}, which is to
find a minimum-weight matching of an auxiliary bipartite graph
$B(\mathbf{x})$, we can convert a feasible solution $\mathbf{x}$ for
the relaxation of (6) to a feasible solution $\mathbf{\hat{x}}$ for
(6). An important observation is that $\mathbf{\hat{x}}$ satisfies
$MS(\mathbf{\hat{x}})\leq 2MS$ and
$E(\mathbf{\hat{x}})=E(\mathbf{x})\leq E(\mathbf{x}^*)$.

 Note that the running time of Shmoys \& Tardos's rounding method \cite{Shmoys93} is
dependent
 on the number of tasks, which is very large in reality \cite{Tarplee14}.  To reduce the running time,
 we will replace  minimum-weight matching by minimum-weight b-matching \cite{Huang11} to design an algorithm
 whose running time is dependent
 on the number of task types. For
completeness, we present the modified Shmoys \& Tardos's rounding
method in \cite{Shmoys93} as follows. Here, for simplicity, we only
show how to construct the bipartite graph $B(\mathbf{x})$ and the
edge weights, ignoring the fraction solution of the matching. Given
a feasible solution $\mathbf{x}$ for the relaxation of (6), let
$x'_{ij}=x_{ij}-\lfloor x_{ij}\rfloor$, for $i=1,\ldots,T$ and
$j=1,\dots,M$. Construct a weighted bipartite graph
$B(\mathbf{x})=(U,V,E;w)$, where  $U=\{u_1,\ldots,u_T\}$ represent
the set of task types. The other node set $V=\{v_{js}| j=1,\ldots,M,
s=1,\ldots,k_j\}$ consists of {\it machine-type} nodes, where
$k_j=\lceil\sum_{i=1}^{T}x'_{ij}\rceil$ and $k_j$ nodes $v_{js}$,
$s=1,\ldots,k_j$, correspond to machine type $j$, for
$j=1,\ldots,M$.

As in \cite{Shmoys93}, the edges in $E$ of the bipartite graph
$B(\mathbf{x})$ will correspond to task-machine pairs $(i,j)$, such
that $x'_{ij}>0$. To construct the edges incident to the nodes in
$V$ corresponding to machine type $j$, sort the task types in order
of nonincreasing estimated times to compute $ETC_{ij}$. For
simplicity, assume that
\begin{eqnarray}
ETC_{1j}\geq ETC_{2j}\geq \ldots\geq ETC_{Tj}. \end{eqnarray}

If $\sum_{i=1}^{T}x'_{ij}\leq 1$, then $k_j=1$, which implies that
there is only one node $v_{j1}\in V$ corresponding to machine type
$j$. For each $x'_{ij}>0$, include $(v_{j1},u_i)\in E$. Otherwise,
find the minimum index $i_1$ such that $\sum_{i=1}^{i_1}x'_{ij}\geq
1$. Let $E$ contain those edges $(v_{j1},u_i)\in E$, $i=1,\ldots,
i_1$, for which $x_{ij}>0$. For each $s=2,\ldots,k_j-1$, find the
minimum index $i_s$ such that $\sum_{i=1}^{i_s}x'_{ij}\geq s$. Let
$E$ contain those edges $(v_{js},u_i)$, $i=i_{s-1}+1,\ldots,i_{s}$,
for which $x'_{ij}>0$. If $\sum_{i=1}^{i_s}x'_{ij}> s$, then also
put edge $(v_{j,s+1},u_{i_s})\in E$. Finally, put edges
$(v_{jk_j},u_i)\in E$, $i=i_{k_j-1}+1,\ldots,T$, for which
$x'_{ij}>0$.

For each edge $(v_{js},u_i)\in E$, let the weight of edge
$(v_{js},u_i)$ be $w(v_{js},u_i)=APC_{ij}ETC_{ij}$. For each
task-type node $u_i\in U$, let the capacity of $u_i$ be
$b_i=\sum_{j=1}^{M}x'_{ij}$, where $b_i$ is an integer as
$\sum_{j=1}^{M}x'_{ij}=\sum_{j=1}^{M}x_{ij}-\sum_{j=1}^{M}\lfloor
x_{ij}\rfloor=T_i-\sum_{j=1}^{M}\lfloor x_{ij}\rfloor$ is an
integer. From the construction of the bipartite graph
$B(\mathbf{x})$, it is easy to verify that there are at most $T$
nodes in $U$ and at most $\sum_{j=1}^{M}k_j\leq MT$ nodes in $V$. As
there are $T+M$ nontrivial constraints in (6), the number of
positive variables in $\mathbf{x}$ is at most $T+M$, following from
the property of linear programming. Combining the fact that there
are one or two corresponding edges in $E$ for each $x'_{ij}>0$,
there are at most $2(T+M)$ edges in $E$. Therefore, the minimum-cost
b-matching ${\cal BM}$, that exactly matches  $b_i$ times  of the
task-type node $u_i$ in $E(\mathbf{x})$, can be found by using the
method in \cite{Huang11}, whose running time is polynomial in $T$
and $M$.

The modified Shmoys \& Tardos's rounding method algorithm to
construct a schedule $x_{ij}$ from a feasible solution $x$ of the
relaxation of (6) is summarized as follows.

 {\sc Algorithm  A}

 {\it Step 1}. Form the bipartite graph $B(\mathbf{x})$ with weights
 on its edges as described above.

 {\it Step 2}.  Use the
method in \cite{Huang11} to find a minimum-weight (integer)
$b$-matching ${\cal BM}$ that
 exactly matches   $b_i$ times of the task-type node   $u_i$ in
 $B(\mathbf{x})$.

 {\it Step 3}. For each edge $(v_{js},u_i)\in   {\cal BM} $, assign a
 task of type $i$ on a machine of type $j$, which implies that
$\hat{x}_{ij}=\lfloor x_{ij}\rfloor+|\{(v_{js},u_i)|(v_{js},u_i)\in
{\cal BM} \}|$, for every $i,j$.

 \vspace{1mm}\noindent {\bf Theorem~2.} \cite{Shmoys93} {\em The schedule
$\mathbf{\hat{x}}$ obtained by {\sc Algorithm  A} has makespan at
most $2MS$, and the energy consumption is at most solution
$E(\mathbf{x}^*)$.}
\subsection{Task-Type-Based Local Assignment}
Recall that a feasible schedule is to assign every indivisible task
to a specific machine. The solution $\hat{x}_{ij}$ obtained in the
last subsection is to assign $\hat{x}_{ij}$ tasks of type $i$ to
machines of type $j$. To obtain a feasible schedule, we need to
schedule the tasks already assigned to each machine type to specific
machines within that group. In a group of machines of type $j$,
$ETC_{ij}$ and $APC_{ij}$ are only dependent on the task type $i$.
Thus, the total energy consumed by machines of type $j$   is
$\sum_{i=1}^{T}\hat{x}_{ij}APC_{ij}ETC_{ij}$, which is a constant.
Therefore, we only need to schedule tasks to minimize makespan,
which is equivalent to the multiprocessor scheduling problem
\cite{Graham69}. Tarplee et al. \cite{Tarplee14} use the common {\it
longest processing time} (LPT) algorithm to assign tasks to machines
for each machine type, where the $\sum_{i=1}^{T}\hat{x}_{ij}$ tasks
are sorted in descending order by execution time, and each task is
assigned to the machine that will complete earliest.

As shown in \cite{Tarplee14}, the effect of the sub-optimality of
LPT algorithm on the overall performance of the systems consider is
insignificant, as the number of tasks is  large generally. However,
this leads to another problem, that the running time of LPT
algorithm will increase dramatically when the number of tasks grows
rapidly. Note that in the HPC system, the number of types of tasks
 is always much less than that of tasks. For example, in the
simulations of \cite{Tarplee14}, there are  30 task types, yet there
are 11,000 tasks. An important observation is that we do not need to
assign one task at a time when assign the tasks of same type. ¡¡

¡¡ Each group of machines of type $j$ is processed independently.
The task types are sorted in descending order by execution time
$ETC_{ij}$, which can be done within $O(T\log T)$ time. Without loss
of generality, assume $ETC_{1j}\geq \cdots \geq ETC_{Tj}$. For each
machine $j_k\in {\cal M}_j$, let $L^i_k$ be the current {\it load}
of machine $j_k$ after assigning tasks of type $i$, where the load
of machine $j_k$ is the total processing time of tasks assigned to
it. Initially, $L^0_k=0$ for each $j_k\in {\cal M}_j$. Let $AL_i$ be
the average load   of machines of type $j$ after assigning the tasks
of type $i$, where
\begin{eqnarray}
AL_i=\frac{\sum_{k: j_k\in {\cal
M}_j}L^{i-1}_k+ETC_{ij}\hat{x}_{ij}}{M_j}.
\end{eqnarray}
 For $k=1$, $\ldots, M_j$, assuming there are $N_{unassign}$ unassigned tasks,  schedule  $\min \{N_{unassign}, N^i_k\}$ tasks of
type $i$ simultaneously to machine $j_k$, where
\begin{eqnarray}
N^i_k=\max \{\lfloor\frac{AL_i-L^{i-1}_k}{ETC_{ij}}\rfloor, 0\}.
\end{eqnarray}
If the load of a machine $j_k$ is increased meaning $N^i_k>0$, we
have
\begin{eqnarray}
AL_i-ETC_{ij}<L^{i}_k=L^{i-1}_k+N^i_k ETC_{ij}\leq
AL_i.\end{eqnarray} Obviously, there are at most $M_j$ unassigned
tasks of type $i$, which can be assigned using LPT algorithm. It is
easy to verify that our method is equivalent to the LPT algorithm in
\cite{Tarplee14}. However, the running time is reduced
 to $O(\sum_{j=1}^{M}(T\log T+TM_j))$, not depending on the number of tasks,
 which is always a huge number in the HPC system.

{\sc Algorithm B}  shows the pseudo-code for assigning tasks to
machines for each type.

\noindent\hrulefill\\
{\sc Algorithm B}  Assign tasks to machines for each type.\\
\vspace*{0.5mm}
 \noindent\hrulefill\\
1:\hspace{1mm} \textbf{For} $j=1$ to $M$ do\\
2:  \hspace{3mm}  Relabel the indices such that
$ETC_{1j}\geq \cdots \geq ETC_{Tj}$;\\
3:  \hspace{3mm}  \textbf{For} $i=1$ to $T$ \textbf{do}\\
4:  \hspace{8mm}   \textbf{For}  each  machine $j_k\in {\cal M}_j$
\textbf{do}\\
5:  \hspace{12mm} Assign $N^i_k$ (defined in (9))  tasks of type $i$
to

 \hspace{8mm} it, if there are  unassigned tasks;\\
6:   \hspace{8mm} \textbf{End for}\\
 7:  \hspace{8mm}  Use LPT
algorithm to assign the remaining tasks

\hspace{8mm} of type $i$ (at most $M_j$);\\
8: \hspace{3mm} \textbf{End for} \\
9:  \hspace{1mm} \textbf{End for}

 \vspace*{1mm}
\noindent\hrulefill\\

 \subsection{Performance Analysis}
In summary, for each  $t\in \{1,\ldots,\lceil\log_{(1+\epsilon)}
 UB/LB\rceil\}$, let $MS=LB(1+\epsilon)^t$. Then, use {\sc Algorithm A} and  {\sc Algorithm B}
to find a feasible solution for (4). Among these solutions (at most
$\lceil\log_{(1+\epsilon)}
 UB/LB\rceil$), choose the one with maximum profit per unit time. It
is easy to verify that the total running time is independent of the
number of tasks.

 For a  maximization problem, if algorithm ${\cal A}$ can produce a feasible solution
with the objective value at least $OPT/\rho$ for any instance, where
$OPT$ denotes the optimal value, then  $\rho$ is called the
worst-case performance ratio or approximation ratio.

Combining (5) and  Theorem 2, the objective of the schedule
$\mathbf{\hat{x}}$ is at least
\begin{eqnarray*}
\frac{p-cE(\mathbf{\hat{x}})}{2MS}&\geq&
\frac{p-cE(\mathbf{x}^*)}{2MS}
 \geq \frac{p-cE(\mathbf{x}^*)}{ 2(1+\epsilon)
MS({\mathbf{x}^*})}\\&\geq &\frac{1}{2+2\epsilon}OPT.
\end{eqnarray*}
It implies that the worst-case performance ratio of the proposed
algorithm is $2+2\epsilon$, for any $\epsilon>0$.
  \section{Experimental Results}
Simulation experiments were performed to compare the quality of TMS
and TTB methods. As in \cite{Tarplee14}, the software was written in
C++ and the LP solver used the simplex method  from COIN-OR CLP
\cite{Clp}.

Without loss of generality, assume that $c=1$ for all the
experiments. As in \cite{Tarplee14}, let $E_{min}$ be the lower
bound on the minimum energy consumed when ignoring makespan, and
$p=\gamma E_{min}$, where $\gamma=p/E_{min}$ is a parameter that
will be used to affect the price per bag. Clearly, when $\gamma$ is
large enough, the focus is to minimize the makespan
\cite{Tarplee14}.  Thus, we only consider the case that $\gamma\in
[1,1.5]$.

For all the simulations, there are nine machine  types and 40
machines of each type for a total of 360 machine, as in
\cite{Tarplee14}. Our first experiment is based on a benchmark
\cite{Benchmark} with nine machine types and five task types,  where
the missing values are deleted. The workload consists of 12, 000
tasks divided among 5 task types. When  $\gamma$ is varying,
different solutions produced by the TMS and TTB methods are shown in
Table 1. The table shows that every solution produced by the TTB
method is better than that produced by the TMS method. Especially,
when $\gamma=1$, because the rounding method in the TMS method will
increase the energy consumption, the TMS method produces a solution
with negative objective value, while the TTB method produces the
optimal solution.

\begin{center}
\begin{tikzpicture}[scale=1.2]
\foreach \x in {1,1.5,2,2.5} \draw (1,\x)--(8,\x); \foreach \x in
{1,2,3,4,5,6,7,8} \draw (\x,1)--(\x,2.5);
 \draw   (1.5,2.25) node {$\gamma=$};
 \draw   (2.5,2.25) node {$1$};
\draw   (3.5,2.25) node {$1.1$}; \draw   (4.5,2.25) node {$1.2$};
\draw (5.5,2.25) node {$1.3$}; \draw   (6.5,2.25) node {$1.4$};
\draw (7.5,2.25) node {$1.5$};

 \draw   (1.5,1.75) node {TMS};
\draw   (1.5,1.25) node {TTB};

\draw (2.5,1.75) node {\small -0.6};
 \draw (2.5,1.25) node
{\small  0.0};

\draw (3.5,1.75) node {\small  985.1};
 \draw (3.5,1.25) node
{\small  986.1};

\draw (4.5,1.75) node {\small  1998.8};
 \draw (4.5,1.25) node
{\small  2009.0};

\draw (5.5,1.75) node {\small  3505.4 };
 \draw (5.5,1.25) node
{\small  3529.8};

\draw (6.5,1.75) node {\small  5491.4};
 \draw (6.5,1.25) node
{\small  5510.7};

\draw (7.5,1.75) node {\small  7933.9};
 \draw (7.5,1.25) node
{\small  7986.0};

 \draw   (4.5,0.5) node {\textbf{Table 1.} The solutions
 with  $\gamma$ varying from 1 to 1.5 };
\end{tikzpicture}
\end{center}

¡¡

Since ¡¡ $ETC_{ij}$ and $APC_{ij}$ differ slightly in the benchmark
\cite{Benchmark}, to quantify the quality of the solutions in a more
general case, we did 25 experiments where $ETC_{ij}$ and $APC_{ij}$
are random numbers between 0 and 1.  In the $q$-th experiment,
$q=1,\ldots,25$, the workload consists of $150q$ tasks divided among
$30$ task types. Figure 2 shows the profit per unit time computed
from the TMS and TTB methods when $\gamma=1.2$. The figure shows
that every solution produced by the TTB method  has a higher profit
per unit time. When the number of tasks is large enough, the
solutions produced by two methods are close to each other.

In fact, for every experiment where $\gamma$ is also a random number
we have done, the TTB method produces a higher quality
solution.¡¡Moreover,  in (6), letting $MS$ be the makespan of the
solution produced by the TMS method, we can obtain a better solution
by using the $b$-matching-based rounding and task-type-based local
assignment method in Section \Rmnum{3}. It is worth to mentioning
that the  TTB method performs much better when $\gamma$ is small or
the average number of tasks per machine is small.

\section{Discussion  and future work}
With small modifications, our algorithm can be extended to the idle
power consumption or the case where there is upper bound on the
allowed power consumption, which are considered in \cite{Tarplee14}.
Due to space constraint, we omit the details here.
\begin{center}
\begin{tikzpicture}[scale=0.8]
\draw (1,1)--(6,1); \draw(1,6)--(6,6);
 \draw[->](1,1)--(1,6);
 \draw(6,1)--(6,6);

\foreach \x in {2,3,4,5,6}
\draw (\x,1)--(\x,1.2);
\foreach \x in {1.5,2,2.5,3,3.5,4,4.5,5,5.5,6}
\draw (1,\x)--(1.05,\x);
\draw (0.7,1) node { 0}; \draw (0.7,1.5) node {10};
 \draw (0.7,2) node {20};
 \draw (0.7,2.5) node {30};
 \draw (0.7,3) node {40};
 \draw (0.7,3.5) node {50};
 \draw (0.7,4) node {60};
 \draw (0.7,4.5) node {70};
 \draw (0.7,5) node {80};
 \draw (0.7,5.5) node {90};
 \draw (0.6,6) node {100};
 \draw   (1.2,1.55) circle (0.5mm);
\filldraw   (1.2,1.76) circle (0.5mm);
 \draw   (1.4,2.05) circle (0.5mm);
\filldraw   (1.4,2.15) circle (0.5mm);
 \draw   (1.6,2.65) circle (0.5mm);
\filldraw   (1.6,2.85) circle (0.5mm);
 \draw   (1.8,3.16) circle (0.5mm);
\filldraw   (1.8,3.3) circle (0.5mm);
 \draw   (2,3.7) circle (0.5mm);
\filldraw   (2,3.8) circle (0.5mm);
 \draw   (2.2,3.69) circle (0.5mm);
\filldraw   (2.2,3.79) circle (0.5mm);
 \draw   (2.4,3.65) circle (0.5mm);
\filldraw   (2.4,3.75) circle (0.5mm);
 \draw   (2.6,4.02) circle (0.5mm);
\filldraw   (2.6,4.15) circle (0.5mm);
 \draw   (2.8,3.75) circle (0.5mm);
\filldraw   (2.8,3.85) circle (0.5mm);
 \draw   (3,4.3) circle (0.5mm);
\filldraw   (3,4.4) circle (0.5mm);
 \draw   (3.2,4.4) circle (0.5mm);
\filldraw   (3.2,4.55) circle (0.5mm);
 \draw   (3.4,4.35) circle (0.5mm);
\filldraw   (3.4,4.5) circle (0.5mm);
 \draw   (3.6,4.7) circle (0.5mm);
\filldraw   (3.6,4.75) circle (0.5mm);
 \draw   (3.8,4.9) circle (0.5mm);
\filldraw   (3.8,4.95) circle (0.5mm);
 \draw   (4,5.08) circle (0.5mm);
\filldraw   (4,5.15) circle (0.5mm); 
 \draw   (4.2,5.07) circle (0.5mm);
\filldraw   (4.2,5.12) circle (0.5mm);
 \draw   (4.4,4.35) circle (0.5mm);
\filldraw   (4.4,4.42) circle (0.5mm);
 \draw   (4.6,4.92) circle (0.5mm);
\filldraw   (4.6,4.95) circle (0.5mm);
 \draw   (4.8,5.1) circle (0.5mm);
\filldraw   (4.8,5.2) circle (0.5mm);
 \draw   (5,5.25) circle (0.5mm);
\filldraw   (5,5.35) circle (0.5mm);
 \draw   (5.2,5.3) circle (0.5mm);
\filldraw   (5.2,5.35) circle (0.5mm);
 \draw   (5.4,5.05) circle (0.5mm);
\filldraw   (5.4,5.52) circle (0.5mm);
 \draw   (5.6,5.2) circle (0.5mm);
\filldraw   (5.6,5.3) circle (0.5mm);
 \draw   (5.8,5.3) circle (0.5mm);
\filldraw   (5.8,5.45) circle (0.5mm);
 \draw   (6,5.1) circle (0.5mm);
\filldraw   (6,5.15) circle (0.5mm);

\filldraw  (6.8,4.5) circle (0.5mm);
 \draw (8,4.5) node {\small TTB method};

\draw   (6.8,3.5) circle (0.5mm); \draw (8,3.5) node {\small TMS
method}; 
 \foreach \x in {1.2,1.4,1.6,1.8}
\draw (\x,1)--(\x,1.1);
 \foreach \x in {2.2,2.4,2.6,2.8}
\draw (\x,1)--(\x,1.1);
 \foreach \x in {3.2,3.4,3.6,3.8}
\draw (\x,1)--(\x,1.1);
 \foreach \x in {4.2,4.4,4.6,4.8}
\draw (\x,1)--(\x,1.1);
 \foreach \x in {5.2,5.4,5.6,5.8}
\draw (\x,1)--(\x,1.1);

 \draw (2,0.7) node {\small 5};
 \draw (3,0.7) node {\small 10};
 \draw (4,0.7) node {\small 15};
 \draw (5,0.7) node {\small 20};
 \draw (6,0.7) node {\small 25};

 \draw (0,3.5)  node {\rotatebox{90}{\small profit per unit time}};

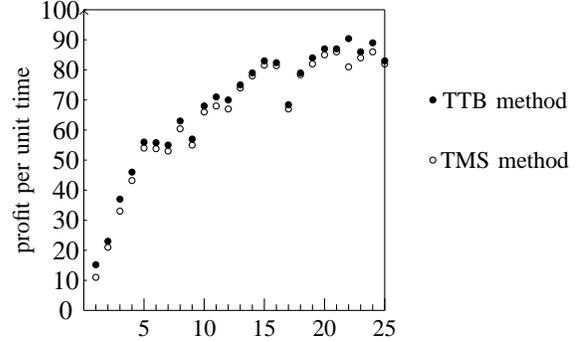
 \draw [xshift=2cm] (3,0) node {\textbf{Figure 3.} 25 randomized
experiments};
\end{tikzpicture}
\end{center}

 Although
experiments show that the solution produced by the TTB algorithms is
close to the optimal solution, this does not hold in a worst-case
scenario. It is interesting and challenging to design a
polynomial-time algorithm with worst-case performance ratio less
than 2.


 \section*{Acknowledgement}
 The work is supported in part by the National Natural Science
Foundation of China [Nos. 11301466, 61170222], and the Natural
Science Foundation of Yunnan Province of China [No. 2014FB114].




\begin{thebibliography}{1}
\bibitem{Tarplee14} K.M. Tarplee, A.A. Maciejewski, and H.J. Siegel,
 ``Energy-aware profit maximizing scheduling algorithm for
heterogeneous computing systems,'' in 14th IEEE/ACM International
Symposium on Cluster, Cloud and Grid Computing, 2014, pp. 595-603.


 \bibitem{Tarplee13} K.M. Tarplee, R. Friese, A.A. Maciejewski, and H.J. Siegel,
 ``Efficient and scalable computation of the energy and makespan
 pareto front for heterogeneous computing systems,'' in Federated
 Conference on Computer Science and Information Systems, Workshop on
 Computational Optimization, 2013, pp. 401-408.




   \bibitem{Shmoys93} D.B. Shmoys, and E. Tardos,  ``An approximation algorithm
for the generalized  assignment  problem,'' Mathematical Programming
62(1-3) (1993), pp. 461-74.

\bibitem{Huang11} B. C. Huang and T. Jebara, ``Fast b-matching via sufficient selection
belief propagation,''  Journal of Machine Learning Research -
Proceedings Track 15 (2011), pp. 361-369.

\bibitem{Graham69} R.
Graham, ``Bounds on multiprocessing timing anomalies,'' SIAM Journal
on Applied Mathematics 17(2) (1969), pp. 416-429.




\bibitem{Clp}(2013, March) Coin-or clp. [Online]. Available:\\ https://projects.
coinor. org/Clp

   \bibitem{Benchmark} (2013, May) Intel core i7 3770k power consumption, thermal.
[Online]. Available: http://openbenchmarking.org/
result/

1204229-SU-CPUMONITO81








\end{thebibliography}
%

\end{document}